\documentclass[reprint, amsmath, amssymb, aps, prl]{revtex4-2} 

\usepackage{amsmath}
\usepackage{amssymb}
\usepackage[cjk]{kotex}
\usepackage{graphicx}
\usepackage{xcolor}
\usepackage{hyperref}

\begin{document}

\title{Interstitial-induced ferromagnetism in a two-dimensional Wigner crystal}

\author{Kyung-Su Kim (김경수)}
 \altaffiliation[]{Corresponding author. \\
 kyungsu@stanford.edu}
\author{Chaitanya Murthy}
\author{Akshat Pandey}
\author{Steven A.~Kivelson}
\affiliation{Department of Physics, Stanford University, Stanford, CA 93405}

\date{\today}

\begin{abstract}
The two-dimensional Wigner crystal (WC) occurs in the strongly interacting regime ($r_s \gg 1$) of the two-dimensional electron gas (2DEG). 
The magnetism of a pure WC is determined by tunneling processes that induce multi-spin ring-exchange interactions, resulting in fully polarized ferromagnetism for large enough $r_s$.
Recently, Hossain et al.~[PNAS 117 (51) 32244-32250] reported the occurrence of a fully polarized ferromagnetic insulator at $r_s \gtrsim 35$ in an AlAs quantum well, but at temperatures orders of magnitude larger than the predicted exchange energies for the pure WC.
Here, we analyze the large $r_s$ dynamics of an interstitial defect in the WC, and show that it produces local ferromagnetism with much higher energy scales.
Three hopping processes are dominant, which favor a large, fully polarized ferromagnetic polaron.
Based on the above results, we speculate concerning the phenomenology of the magnetism near the metal-insulator transition of the 2DEG.
\end{abstract}


\maketitle

The two-dimensional electron gas (2DEG) has proven to be a rich platform for studying strongly correlated phases of matter, despite its deceptively simple Hamiltonian
\begin{align}
    \label{eq:H}
    H = \sum_{i}\frac{\vec p_i^{\, 2}}{2m} + \sum_{i<j} \frac{e^2}{4\pi \epsilon} \frac{1}{|\vec r_i -\vec r_j|} .
\end{align}
The important dimensionless parameter in the problem is the ratio $r_s$ of the typical interaction and kinetic energies; $r_s = 1/(a_\mathrm{B} \sqrt{\pi n})$, where $n$ is the electron density and $a_\mathrm{B} = 4\pi \epsilon \hbar^2 /m e^2$ is the effective Bohr radius.
The electrons form an unpolarized Fermi liquid (FL) when $r_s$ is small, whereas 
a Wigner crystal (WC) phase occurs when $r_s > r_{\mathrm{wc}} \approx 31 \pm 1$~\cite{Wigner1934, tanatar1989QMC, drummond2009QMC, attaccalite2002QMC}.
Recently, experiments on ``ultraclean'' AlAs quantum wells reported the appearance of a fully polarized ferromagnetic insulating phase when $r_s \gtrsim 35$~\cite{hossain2020ferromagnetism, hossain2021valley, kim2021discovery}, where the WC physics may play a key role.
Ferromagnetic tendencies near the metal--insulator transition have also been seen in a variety of other 2DEG systems~\cite{ferro1, ferro2}.
In this paper we explore a new mechanism of ferromagnetism in the large-$r_s$ regime.

There have been many previous theoretical studies of the magnetism of the WC~\cite{Roger1984WKB, chakravarty1999WC, Katano2000WKB,  Ceperley2001exchange, voelker2001disorder}. 
Deep within the WC phase (in the $r_s \to \infty $ limit), a semi-classical instanton method allows an asymptotically exact calculation of various multi-spin ring exchange energies $J_{\mathrm{wc}}$.
The result of these calculations is that the WC (and hence the 2DEG) is fully spin-polarized in the $r_s \to \infty$ limit~\cite{Roger1984WKB, chakravarty1999WC, Katano2000WKB}. 
This result has been corroborated by a path integral Monte Carlo calculation~\cite{Ceperley2001exchange}.
Therefore it is tempting to say that the observed fully polarized ferromagnetic insulator is the ferromagnetic WC.
However, we will see that such a mechanism provides a minuscule energy scale (i.e.~temperature scale $T^*$) for the ferromagnetism, which is much below those accessed in the experiments.
Moreover, the theoretical studies suggest~\cite{Roger1984WKB, chakravarty1999WC, Katano2000WKB, Ceperley2001exchange, voelker2001disorder} that the dominant exchange interactions are actually antiferromagnetic in the experimentally relevant range of $r_s\sim 40$ of the 2DEG.

\begin{figure}[t]
    \centering
    \includegraphics[width = 0.5 \textwidth]{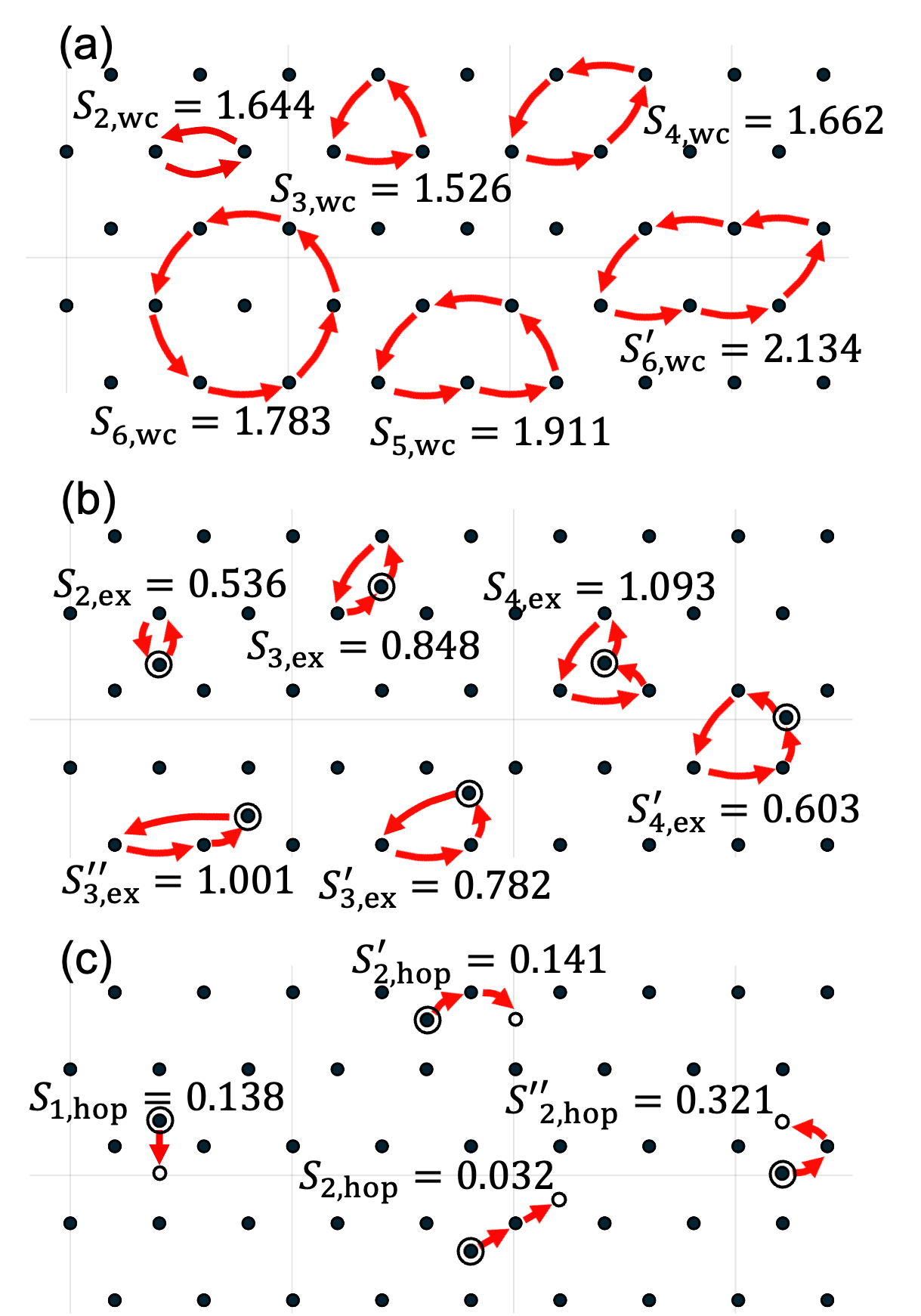}
    \caption{
    Schematic of various exchange and hopping processes along with the corresponding dimensionless actions $S_a$: 
    (a) Exchange processes in the pure WC.
    (b) New exchange processes in the WC induced by a triangle-centered interstitial.
    (c) Hopping processes in the WC induced by the interstitial.
    In panels b \& c, dots surrounded by a circle denote initially occupied interstitial sites, while open circles denote final interstitial sites which are initially vacant.
    The dimensionless actions in the panel a are quoted from Ref.~\cite{voelker2001disorder}.
    Panels b \& c show the main results of this paper, calculated with a system of size $10\times 12 +1$ starting from the {\it relaxed} triangle-centered interstitial configuration (see the main text and the Supplementary Material for details). 
    The corresponding values for the exchange couplings, $J_a$, and the hopping matrix elements, $t_a,$ are then computed using Eq.~\eqref{eq:tunneling} and its analogue.
    }
    \label{fig:summary}
\end{figure}

We instead propose a new mechanism for ferromagnetism at large $r_s$, induced by interstitial defects centered at triangular plaquettes of the WC~\cite{Cockayne1991defect, Fisher1979defect, candido2001defect}.
(This idea was inspired by a related, but distinct, earlier proposal by Spivak and collaborators~\cite{meierovichSpivak1981, spivak2000ferromagnetism} of ferromagnetism produced by interfacial fluctuations between a WC and a FL.)
The presence of interstitials generates additional exchange ($J_a$) and hopping ($t_a$) processes, which we calculate using the semi-classical instanton method. 
See Fig.~\ref{fig:summary} for a summary of the results.
Three hopping processes turn out to have (exponentially) large energy scales compared to any exchange energy of the defect-free WC.
We prove that a single interstitial fully polarizes a large region of the WC (i.e.~produces a large ferromagnetic polaron), and argue that a dilute concentration of interstitials will lead to a fully polarized ferromagnetic ground state.
Moreover, the characteristic temperature scale of the ferromagnet is $T^* \sim \nu_{\mathrm{int}} \cdot t $, where $0 \leq \nu_{\mathrm{int}} \leq 1$ is the filling of interstitial sites and $t$ is an appropriate sum of hopping energies $t_a$.
At the values of $r_s$ pertinent to the experiments, $T^*$ is in the experimentally relevant range, even for a low concentration of interstitials.

On the more phenomenological level, near the metal-insulator transition, it is likely that the 2DEG forms a spatially inhomogeneous mixture of regions that exhibit local WC order (with slightly lower than average electron density) coexisting with puddles of FL (with slightly higher density).
This can arise as a consequence of disorder~\cite{imry1979influence, aizenman1989rounding} or could reflect the electronic micro-emulsion phases expected when macroscopic phase separation is frustrated by long-range interactions~\cite{spivak2003phase, spivak2004phases, jamei2005universal, spivak2006transport}.
Consequently, a finite density of extra electrons will be induced at the boundaries of WC regions. 
The lowest energy defect that can accommodate an extra electron is known to be the triangle-centered interstitial ~\cite{Cockayne1991defect, Fisher1979defect,candido2001defect}.

\vspace{0.5em}
{\bf Semi-classical derivation of the effective Hamiltonian.}
For orientation, we start by recapitulating the semi-classical theory of magnetism in the WC.
In the $r_s \to \infty$ limit, the Coulomb interaction dominates, and the electrons form a WC~\cite{Wigner1934} with all spin states degenerate.
The kinetic energy lifts this degeneracy by inducing virtual tunneling processes among WC sites.
The effective spin Hamiltonian can be written as a sum over ring exchange terms:
\begin{align}
    H_{\mathrm{eff}}^{\mathrm{wc}} = \sum_{a} (-1)^{n_a}\, {J_a} \, ( \mathcal{\hat P}_a +\mathcal{\hat P}^{-1}_a ).
\end{align}
Here, $a=(i_1,i_2,\dots,i_{n_a})$ labels a ring exchange process involving $n_a$ sites, $i_1 \rightarrow i_2 \rightarrow \dots \rightarrow i_{n_a} \rightarrow i_1$, 
and $\hat{P}_a$ is the corresponding $n_a$-particle cyclic permutation operator.
$\hat{P}_a$ can, in turn, be expressed as a product of two-particle exchange operators, each of which can be written in terms of spin operators as $\hat{P}_{(i,j)} = 2(\vec{S}_i \cdot \vec{S}_j + \frac{1}{4})$.
All exchange couplings $J_a$ are positive; 
the signs $(-1)^{n_a}$ are fixed by anti-symmetry of the many-body wave-function, which implies that exchanges involving an even (odd) number of electrons are antiferromagnetic (ferromagnetic)~\cite{thouless1965}.
The exchange energies $J_a$ can be calculated using the semi-classical instanton method, which is asymptotically exact in the $r_s \to \infty $ limit:
\begin{align}
\label{eq:tunneling}
    J_a = \hbar \omega_0
    \left( \frac{\sqrt{r_s} \, S_a}{2\pi} \right)^{1/2} A_a \, 
    \exp\!\big[- \sqrt{r_s} \, S_{a} \big] .
\end{align}
Here, $\hbar\sqrt{r_s} \, S_a$ is the classical Euclidean action along the minimal action path that implements the particle exchange labelled by $a$, 
$S_a$ is the ``dimensionless action,'' which is independent of $r_s$,
and $\hbar\omega_0/2 = {1.6274}/{r_s^{3/2}}$ is the zero-point phonon energy (per particle) of the defect-free WC in units of the effective Rydberg energy ${\rm Ry }= e^2/8\pi\epsilon a_{\mathrm{B}}$~\cite{Cockayne1991defect, bonsall1977some}.
$A_a$ is the dimensionless magnitude of the fluctuation determinant~\cite{altlandSimons2010, coleman1988aspects}, which is generally of order 1.
Including all $r_s$ dependencies, $J_a = O(r_s^{-5/4} e^{- \sqrt{r_s} \, S_a})$.
To simplify notation, we often suppress the full indices $a$ in the subscripts of $J_a$ and $S_a$, and instead label these by $n_a$---if there are multiple processes involving the same number of particles, we distinguish them with primes (e.g. $S_{4,{\rm wc}}$ and $S'_{4,{\rm wc}}$, etc.).

In Fig.~\ref{fig:summary}a, we illustrate the six most important exchange processes for the pure WC and quote the dimensionless actions calculated by Voelker and Chakravarty in Ref.~\cite{voelker2001disorder}. 
Although the dimensionless actions for all these processes are quite comparable, the (ferromagnetic) three-particle ring exchange process has the smallest action and hence determines the magnetism in the $r_s\to \infty$ limit~\cite{Roger1984WKB, chakravarty1999WC, Katano2000WKB, voelker2001disorder}. 
The characteristic temperature scale for ferromagnetism, $T^*$, is set by $J_3$; 
evaluating Eq.~\eqref{eq:tunneling} at $r_s \approx 40$ with the parameters of AlAs and the fluctuation determinant, $A_3 = 2.19$, calculated in Ref.~\cite{voelker2001disorder}, we find $T^* \sim 0.003\,\mathrm{K}$.
This is two orders of magnitude smaller than the temperature at which the experiments are done ($T\gtrsim 0.3\,\mathrm{K}$)~\cite{hossain2020ferromagnetism}.

\begin{figure*}[t]
\centering
    \includegraphics[width=\textwidth]{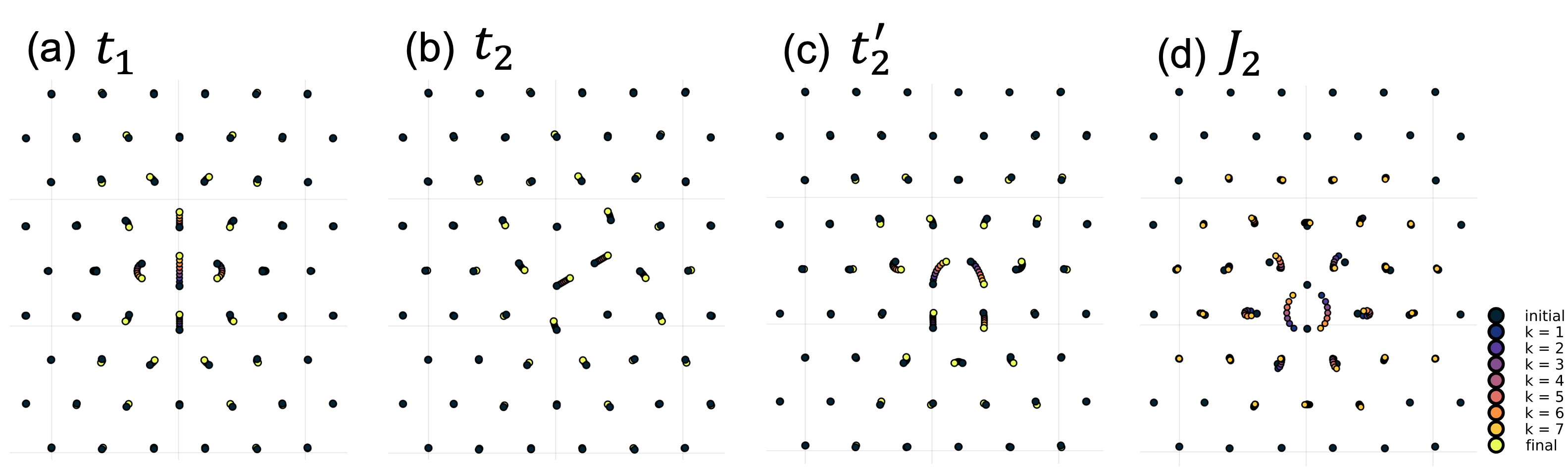}
    \caption{
    Visualization of multiparticle tunneling trajectories involving a (relaxed) triangle-centered interstitial: 
    (a--c) The three dominant interstitial hopping processes.  
    (d) The most important exchange process involving an interstitial. 
    The corresponding dimensionless actions are 
    (a) $S_{1, \mathrm{hop}}= 0.138$, 
    (b) $S_{2, \mathrm{hop}}=0.032$, 
    (c) $S'_{2, \mathrm{hop}}=0.141$ and 
    (d) $S_{2, \mathrm{ex}} = 0.536$, as shown in Fig.~\ref{fig:summary}. 
    The colors indicate seven intermediate configurations, indexed by $k$, along with the initial and the final configuration.
    }
    \label{fig:interstitial tunneling}
\end{figure*}

In the presence of a triangle-centered interstitial in the WC, new tunneling processes are introduced (Figs.~\ref{fig:summary}b \& c).
The semi-classical expression~\eqref{eq:tunneling} can be used to calculate both exchange interactions involving an interstitial, $J_a$, and interstitial hopping processes, $t_a$ (where again $a$ labels a particular process).
The dimensionless action $S_a$ is calculated numerically by minimizing the Euclidean action $\hbar \sqrt {r_s} \, S = \int_{\mathbf{X_i}}^{\mathbf{X_f}} dX \sqrt{2m(V-E_0)}$ on a supercell containing $10\times 12+1$ electrons (including the interstitial) with periodic boundary conditions.
Here, $V$ is the Coulomb interaction, $E_0$ is the energy cost of introducing one interstitial in the WC, and $\mathbf{X_i}$ and $\mathbf{X_f}$ are the initial and the final relaxed interstitial configurations, respectively. 
We discretize the tunneling path to $7$ intermediate configurations and allow up to $30$ electrons to adjust their positions during the minimization. 
For exchange processes, all the remaining electrons are fixed at their initial positions, whereas for hopping processes, they move in linearly interpolating paths connecting the initial and the final positions.
For the minimization, we used the limited-memory Broyden–Fletcher–Goldfarb–Shanno (L-BFGS) algorithm~\cite{optim}.
Coulomb interactions are treated with the standard Ewald method.
See Supplementary Material for more details of the calculations.

Fig.~\ref{fig:summary}b schematically shows various multi-particle exchange processes that involve an interstitial (circled), along with the corresponding dimensionless actions $S_{a,\mathrm{ex}}$.
Interstitial hopping processes are shown schematically in Fig.~\ref{fig:summary}c, along with the dimensionless actions $S_{a,\mathrm{hop}}$.
Among these, one cooperative hopping term, $t_2$, clearly dominates, as its action, $S_{2,\mathrm{hop}} = 0.032$, is more than an order of magnitude smaller than most others. 
(Recall that $\sqrt{r_s} \, S_a$ appears in the exponent of the expressions for $J_a$ or $t_a$!)
However, the $t_2$ term does not connect all the WC sites in the presence of one interstitial, so by itself, it cannot fully lift the ground state spin degeneracy (see Fig. S2 of the Supplementary Material).
The next dominant terms are $t_1$ and $t_2'$ (corresponding to $S_{1,\mathrm{hop}}$ and $S'_{2,\mathrm{hop}}$ in Fig.~\ref{fig:summary}c).
Together with $t_2$, these terms fully determine the magnetism of the WC in the presence of a small density of interstitials.
We visualize the tunneling paths corresponding to these three processes, along with one exchange process, in Fig.~\ref{fig:interstitial tunneling}.
Keeping these three dominant terms results in an effective Hamiltonian:
\begin{align}
\label{eq:H_eff}
    H_{\mathrm{eff}} = 
    &\, -t_2 \sum_{\substack{(n,j,n^\prime) \\ \in (t_2 \: \mathrm{path})}} 
    \sum_{\sigma,\sigma' = \uparrow,\downarrow} 
    c^{\dagger}_{n,\sigma'} f^{\dagger}_{j,\sigma} f_{j,\sigma'} c_{n^\prime,\sigma} \nonumber \\
    &\, -t_2' \sum_{\substack{(n,j,n^\prime) \\ \in (t_2' \: \mathrm{path})}}
    \sum_{\sigma,\sigma' = \uparrow,\downarrow} 
    c^{\dagger}_{n,\sigma'} f^{\dagger}_{j,\sigma} f_{j,\sigma'} c_{n^\prime,\sigma} \nonumber \\
    &\, -t_1 \sum_{\left<n,n^\prime\right>}
    \sum_{\sigma=\uparrow,\downarrow} 
    c^{\dagger}_{n,\sigma}c_{n^\prime,\sigma} \ + \  \left [U=\infty\right ] .
\end{align}
Here, $f_{j\sigma}^\dagger$ is the creation operator of localized electrons that live on the  triangular lattice sites $j$, and $c_{n,\sigma}^\dagger$ is the creation operator of itinerant electrons that live on the triangular plaquette centers $n$.
The last $U=\infty$ condition precludes any doubly occupied sites.
One can check explicitly that all these $t_a$'s are positive.

The remaining tunnelling terms that we have omitted from $H_{\mathrm{eff}}$, including the exchange terms $J_a$, are exponentially smaller than those we have kept.
We have also omitted direct (elastic) interactions between interstitials which are small only in proportion to powers of $1/r_s$. 
These are negligible both because we are interested in the situation with a dilute concentration of interstitials, and because they turn out to be small in the experimentally relevant range of $r_s$~\footnote{%
The elastic interactions are expected to scale as $O(r_s^{-3})$~\cite{Fisher1979defect}, so they are indeed large in the strict asymptotic sense compared to any tunnelling term. 
However, the very small value of, for example, $S_{2,\mathrm{hop}}$ means that even for quite large $r_s$, the exponential factor is not dominant, e.g.~$\exp( -\sqrt{r_s} \, S_{2,\mathrm{hop}}) = 0.64$ at $r_s=200$. 
Thus, the fact that the prefactor of the tunnelling terms is $O(r_s^{-5/4})$ means that those with small action tend to be larger than the elastic interactions in the experimentally relevant range of $r_s$.
}.

\vspace{0.5em}
{\bf A single interstitial.}
In the presence of one interstitial in the WC, we prove the following theorem---reminiscent of the proof of Nagaoka ferromagnetism in the $U=\infty$ Hubbard model---using the Perron-Frobenius theorem~\cite{nagaoka1966ferromagnetism, thouless1965, tasaki1989proof}:

\vspace{0.25em}
\underline{Theorem}: 
The ground state of $H_{\mathrm{eff}}$ in any finite system in the presence of a single interstitial (i.e., for $\nu = N^{-1} \sum_{j,\sigma} f_{j,\sigma}^\dagger f_{j,\sigma} = 1$ and $\sum_{n,\sigma} c_{n,\sigma}^\dagger c_{n,\sigma} = 1$, where $N$ is the number of WC sites $j$) is the fully polarized ferromagnet; it is unique up to global spin rotations.

\vspace{0.25em}
\underline{Proof}: 
$H_{\mathrm{eff}}$ commutes with the total spin operator $\vec{S}_{\mathrm{total}}$, so its spectrum consists of degenerate multiplets with definite $S^2_{\mathrm{total}}$. 
We show that the ground state multiplet has maximal $S^2_{\mathrm{total}}$.
We restrict attention to the sector of Hilbert space with $S^z_{\mathrm{total}} = 0$ for $N+1$ even and  $S^z_{\mathrm{total}} = \frac{1}{2}$ for $N+1$ odd, since these lowest $|S^z_{\mathrm{total}}|$ sectors contain one representative state from each multiplet.
We define basis states
\begin{align}
\label{eq:basis}
    \left | n,\tau,\{\sigma\} \right > 
    \equiv c^{\dagger}_{n, \tau} f^{\dagger}_{1,\sigma_1} \cdots f^{\dagger}_{N,\sigma_N} 
    \left | 0 \right > ,
\end{align}
where $n$ is the position of the interstitial electron, $\tau$ is its spin,
and the $\sigma_j$'s specify the spins of the WC sites, which we number in an arbitrary manner from $j=1$ to $N$.
All the basis states in Eq.~\eqref{eq:basis} can be reached from any starting state by repeated application of the hopping operators in $H_{\mathrm{eff}}$ [Eq.~\eqref{eq:H_eff}]---we say that the hoppings satisfy the ``connectivity condition''~\footnote{%
That the connectivity condition is satisfied can be seen as follows: 
The $t_2$ (or $t_2'$) processes exchange the spin of the interstitial with that of a neighboring WC site, while $t_1$ processes simply move the interstitial around. 
By composing these, we can arbitrarily permute the spins of the WC.
}.

We now consider matrix elements of $H_{\mathrm{eff}}$ in this basis:  
It is easy to see that any state that has a non-zero matrix element with $\left | n, \tau, \{\sigma\} \right >$ must be of the form
\begin{align}
    \left | n', \sigma_j, \{\sigma_1, \cdots, \sigma_{j-1}, \tau, \sigma_{j+1}, \cdots, \sigma_N \} \right > 
    \ \ \mathrm{or} \ \ 
    \left | n', \tau, \{\sigma\} \right > .
    \nonumber
\end{align}
Moreover, it is a simple algebra to show that
\begin{align}
    &\left < n', \sigma_j, \{\sigma_1, \cdots, \sigma_{j-1}, \tau, \sigma_{j+1}, \cdots, \sigma_N\} \right | H_{\mathrm{eff}} \left | n, \tau, \{\sigma\} \right > \nonumber \\  
    &= -t_2 \ \ \mathrm{or} \ -t_2',
\end{align}
and
\begin{align}
    \left < n',\tau,\{\sigma\} \right | H_{\mathrm{eff}} \left | n,\tau,\{\sigma\} \right >
    = -t_1 ,
\end{align}
depending on which of the three hopping terms connect the two states.
Since $H_{\mathrm{eff}}$ satisfies the connectivity condition and all matrix elements are non-positive, the Perron-Frobenius theorem implies that the ground state is unique and is a superposition of all the basis states $\left | n,\tau,\{\sigma\} \right >$ with positive coefficients.
This state is necessarily a maximal spin state, i.e.~has total spin $S_{\mathrm{total}} = (N+1)/2$.
$\square$

\vspace{0.25em}
Note that, in the $S^z_{\mathrm{total}} = (N+1)/2$ sector, $H_{\mathrm{eff}}$ is a non-interacting Hamiltonian, whose ground state is the state where the interstitial electron is in a Bloch state with $\vec{k} = \vec{0}$; 
the state we have found in the minimal $|S^z_{\mathrm{total}}|$ sector is thus related to this state by repeated applications of the global spin-lowering operator.

\vspace{0.5em}
{\bf Phase diagram.}
Although the exchange terms omitted in Eq.~\eqref{eq:H_eff} are exponentially smaller than those we have kept, the former terms can be important when considering the thermodynamic limit, $N \to \infty$.
In particular, whenever the bulk exchange couplings $J_a$ favor anything other than the ferromagnetic state, a single interstitial can only polarize a finite number of WC sites to become a ferromagnetic polaron~\cite{kivelson2022hubbard}.
(Note that a Monte Carlo study found that for the pure WC, antiferromagnetic correlations are favored for $r_s \lesssim 175$~\cite{Ceperley2001exchange}.)
The size of the ferromagnetic polaron is determined by the competition between the energy gain to delocalize the interstitial within a region of radius $R$, $t \cdot (a/R)^2$, and the energy cost, $J \cdot (R/a)^2$, to destroy the antiferromagnetism there, where $J$ is an appropriate sum of the microscopic antiferromagnetic exchange interactions, and $a$ is a lattice constant of the WC.
Minimizing the free energy, we obtain the size of the ferromagnetic polaron:
\begin{align}
    R_{\mathrm{polar}}^2 
    \sim a^2 \, \sqrt{t/J} 
    \sim a^2 \, \exp\!\left( \tfrac{1}{2} \sqrt{r_s} \, \alpha_{\mathrm{polar}} \right) ,
\end{align}
where $t$ is an appropriate sum of $t_2$, $t_2'$ and $t_1$. 
(When $t > T > J$, $J$ is substituted by $T$ in the estimate of the polaron size.)
By comparing the results for $J_a$ and $t_a$ summarized in Fig.~\ref{fig:summary}, it is to be expected that $\alpha_{\mathrm{polar}} \approx 1$.

The properties of $H_{\mathrm{eff}}$ with a finite filling of interstitials, $\nu_{\mathrm{int}} > 0$, are non-trivial, and the complexity is increased if we include the effect of antiferromagnetic interactions, $J > 0$.  
However, for $t/J \gg 1$, certain general features of the phase diagram can be inferred by analogy with the behavior of the ordinary Hubbard model at large $U/t$ in the presence of a dilute concentration of holes~\cite{emeryKivelson1990phaseSeparation, Altshuler2002phaseSeparation, kivelson2022hubbard, liu2012dmrg, spivak2000ferromagnetism}: 
It is likely that at $T=0$, for a range of dopings $\nu_{\mathrm{int}} \in (0, \nu_c)$, there is two-phase coexistence between an insulating antiferromagnetic phase and a half-metallic ferromagnetic phase, with $\nu_c \sim a^2/R_{\mathrm{polar}}^2$. 
The fully polarized ferromagnetic phase then likely appears for a range of fillings, $\nu_{\mathrm{int}} > \nu_c$.
Furthermore, the temperature scale for the onset of ferromagnetism can be estimated to be proportional to the Fermi energy, $T^* \sim \nu_{\mathrm{int}} \cdot t$.

\vspace{0.5em}
{\bf Quantitative considerations in AlAs.}
To flesh out the general discussion, we evaluate various quantities with the parameters relevant to AlAs ($\epsilon = 10 \, \epsilon_0$ and $m = 0.46 \, m_{\mathrm{e}}$, where $\epsilon_0$ is the vacuum permittivity and $m_{\mathrm{e}}$ is the electron mass) in the insulating phase close to the metal-insulator transition, i.e.~with $r_s \approx 40$.
The zero-point phonon energy in the presence of an interstitial is $\hbar \omega_0 / 2 = {1.034}/{r_s^{3/2}}$ in units of the effective Rydberg energy, $\mathrm{Ry} = 731\,\mathrm{K}$~\cite{Cockayne1991defect}.
Using the same value for the fluctuation determinant as for $J_3$ of the pure WC, $A_3=2.19$, Eq.~\eqref{eq:tunneling} gives 
$t_2 \sim 1.9\,\mathrm{K}$, 
$t_2' \sim 2\,\mathrm{K}$, 
$t_1 \sim 2\,\mathrm{K}$, and hence 
$t \sim t_2 + t_2' + t_1 \sim 6\,\mathrm{K}$, 
a much higher energy scale than that of the pure WC for which $J \sim 0.003\,\mathrm{K}$.
(The latter is in the same ball-park as estimates of $J$ from the path integral Monte Carlo calculation~\cite{Ceperley2001exchange}.)
This means that for $\nu_{\mathrm{int}} \approx \nu_c$, the temperature scale for ferromagnetism is $T^* \sim \sqrt{J t} \sim 0.1\,\mathrm{K}$.

\vspace{0.5em}
{\bf Phenomenological considerations.}
While our calculations 
show that an interstitial in a WC generates a large ferromagnetic polaron, the relevance of this observation to any experimental system turns on other considerations.  The existence of a finite concentration of interstitials is surely not a universal feature of a WC phase.

Let us first consider the scenario in which a small density of interstitials are introduced from nearby coexisting (higher-density) Fermi-liquid (FL) regions, as discussed earlier. 
If the interstitial density is sufficiently large, and if the WC regions percolate throughout the sample, it can result in a ferromagnetic phase in which the WC regions are fully polarized. 
Given that the FL at large $r_s$ has a large ferromagnetic susceptibility, it is also possible to imagine circumstances in which the FL puddles, as well, are driven ferromagnetic by their interactions with the ferromagnetic WC.
We propose that such a picture may apply to the fully polarized insulating phase found in AlAs quantum wells~\cite{hossain2020ferromagnetism}.

We can also imagine  cases in which interstitials are induced by extrinsic sources even in the absence of FL regions: e.g.~due to 
a slowly varying disorder potential 
and/or 
a weak commensurate locking of the WC to the  potential from the underlying semiconductor (especially when this period is large, as in a Moir\'e system).
We note that, in contrast to the AlAs system, a fully-polarized insulating phase is not observed in a recent experiment on another 2DEG in a MgZnO/ZnO heterostructure \cite{falson2022competing}.  What material-specific aspect of these systems is responsible for this dichotomy is presently unclear.  These are all issues we hope to address in future work.



%
%
\section{Acknowledgments}
We thank Boris Spivak for initial insights which led to this investigation.
We also thank Peter Littlewood, Eun-Ah Kim, Ilya Esterlis, Mansour Shayegan, Brian Skinner, Inti Sodemann and Joseph Falson for helpful comments on the draft.
This work was supported in part by NSF grant No.~DMR-2000987 at Stanford (KSK and SAK),
the Gordon and Betty Moore Foundation's EPiQS Initiative through GBMF8686 (CM), 
and the Stanford Graduate Fellowship (AP).
Parts of the computing for this project were performed on the Sherlock computing cluster at Stanford University.


\bibliography{apssamp.bib}

\end{document}


\author{Kyung-Su Kim (김경수)}
 \altaffiliation[]{kyungsu@stanford.edu}
\author{Chaitanya Murthy}%
\author{Akshat Pandey}
\author{Steven A. Kivelson}
\affiliation{Department of Physics, Stanford University, Stanford, CA 93405}%

\title{\huge Supplementary material}

\maketitle

\section{1. Details of semiclassical calculations in the $r_s \to \infty $ limit}

The energy splitting $\Delta \epsilon$ (hence the corresponding matrix elements) due to particle exchanges or hopping processes can be calculated using the semi-classical instanton method to yield \cite{altlandSimons2010, coleman1988aspects}:
\begin{align}
    \Delta \epsilon =   A\cdot  \hbar \omega_0\  \sqrt{\frac{S_0}{2\pi \hbar}} e^{-S_0/\hbar},
\end{align}
where $A$ is a fluctuation determinant that we will approximate to be 1, $\hbar \omega_0/2 $ is a zero-point (phonon) energy above a classical ground-state energy, and $S_0$ is a one-instanton action for the exchange/hopping process.
The action in imaginary time is expressed as  
\begin{align}
    S = \int d\tau \sum_{i=1}^N \frac{m}{2}{\bf \dot x}_i^2 + \frac{e^2}{4\pi \epsilon} \sum_{i<j} \frac{1}{|{\bf x_i}-{\bf x_j}|}   \equiv   \int d\tau  \left (T + V \right ) .
\end{align}
We will collectively denote the coordinates of $N$ particles as $\bf X \equiv ({\bf x_1},...,{\bf x_N} )$: this covers a $2N$-dimensional configuration space.  
The instanton action $S_0$ is obtained by solving the classical equation of motion, $m\frac{d^2 {\bf X_{\rm cl}} }{d\tau ^2 } = \nabla V({\bf X_{\rm cl}})$:
\begin{align}
\label{eq:classical action}
    S_{\rm cl}/\hbar = \int_{{\bf X_i}}^{{\bf X_f}} dX \sqrt{2m(V-E_0)}/\hbar = \sqrt{ r_s } \int_{\tilde{\bf X_i}}^{\tilde{\bf X_f}} d\tilde X \sqrt{2(\tilde V-\tilde E_0)} ,
\end{align}
where the integration is over the classical path  ${\bf X_{cl}}(\tau)$, the energy 
\begin{align}
    E_0 \equiv -\frac{1}{2} m (\frac{d{\bf X_{\rm cl}}}{d\tau})^2 +V({\bf X_{\rm cl}}) = V({\bf X_i})
\end{align}
is the Coulomb energy of the classical minima, $\tilde {\bf X} = {\bf X}/a $ is a normalized coordinate, $ a = 1/\sqrt{\pi n}$ is a typical distance between electrons in the WC, and $r_s = \frac{a}{a_B}= \frac{me^2 a}{4\pi \epsilon \hbar^2}$ is the ratio of the typical interaction strength to kinetic energy.
Hence, the WC lattice constant in the normalized coordinate $\tilde {\bf X}$ is $\tilde l_a = \sqrt{\frac{2\pi}{\sqrt 3}}$. Finally, $\tilde V(\tilde X) = \sum_{i<j} \frac 1 {| {\bf \tilde x_i} -  {\bf \tilde x_j}|}$ is the dimensionless Coulomb energy.
We can now work in dimensionless coordinates and action:
\begin{align}
    \tilde S[\tilde{\bf X}] = \int d\tau \left [ \frac{1}{2}{\bf \dot X}^2 +  \sum_{i<j} \frac{1}{|{\bf \tilde x_i}-{\bf \tilde x_j}|}  \right].
\end{align}
(We will drop the tildes henceforth.)
The problem is now reduced to finding the path ${\bf X}(\tau)$ that minimizes the action $S_0 = \int _{{\bf X}_i}^{{\bf X}_f}\sqrt{2\Delta V({\bf X})} \ dX $ for various processes, where 
\begin{align}
    \Delta V(\bf X) \equiv V({\bf X}) - E_0 = \sum_{i<j} \frac{1}{ |\bf x_i - \bf x_j | } - \sum_{i<j} \frac{1}{ |\bf x_i^{(0)} - \bf x_j^{(0)} | }.
\end{align}
Here, $\bf X^{(0)}$ denotes an initial configuration corresponding to a certain classical minimum of $V(\bf X)$. For example, for exchange processes in a pure WC, $\bf X^{(0)}$ is the triangular lattice WC configuration and for interstitial tunneling processes, it is the relaxed interstitial configuration.
\section{2. Ewald method}
The numerical calculation is done using a supercell containing $M=10\times 12 +1 $ electrons (the $+1$ corresponding to one interstitial) with periodic boundary conditions.
In calculating the Coulomb energy $V$, we used the standard Ewald method (see e.g. \cite{Roger1984WKB}).
We briefly summarize the method here:
The interaction energy between an electron at site $\vec r_i$ and another electron at $\vec r_j$ together with its periodic images $\vec r_j + \vec R_l$, where $\vec R_l $ are supercell lattice vectors, can be expressed as 
\begin{align}
\label{eq:Ewald}
    &\sum_{l} \frac{1}{|\vec r - \vec R_l|} = \frac{2}{\sqrt{\pi}} \int_0^\infty du\  \frac{\pi}{s u^2}\sum_{\vec G} {\rm e}^{i \vec G \cdot \vec r} {\rm e}^{-\frac{|\vec G|^2}{4u^2}} \nonumber \\ 
    &= \frac{2}{\sqrt{\pi}} \int_0^ \epsilon du\  \frac{\pi}{s u^2}\sum_{\vec G} {\rm e}^{i \vec G \cdot \vec r} {\rm e}^{-\frac{|G|^2}{4u^2}} + \frac 2 {\sqrt \pi} \sum_l \int_\epsilon ^\infty du {\rm e}^{-u^2|\vec r - \vec R_l|^2}
    \nonumber \\
    &= \frac{2\pi }{s} \sum_{\vec G} \frac 1 {|\vec G|} {\rm erfc}\left (\frac{|\vec G|}{2\epsilon} \right) {\rm e}^{i \vec G \cdot \vec r} + \sum_l \frac 1 {|\vec r - \vec R_l|} {\rm erfc} \left ( \epsilon|\vec r - \vec R_l| \right )
    .
\end{align}
Here $\vec r = \vec r_i -\vec r_j$, $\vec G $ are the reciprocal lattice vectors of the supercell, erfc is the complementary error function, and $s=\frac{\sqrt 3}{2}\tilde l_a^2$ is the area of the unit cell of the pure WC (again, $\tilde l_a = \sqrt{\frac{2\pi}{\sqrt 3}}$).
In the first identity, the integral transfomation $\frac 1 {|\vec r -\vec R_l|} = \frac{2}{\sqrt \pi }\int _0^\infty du\  {\rm e}^{-u^2|\vec r - \vec R_l|^2}$ and the Poisson summation formula were used.
In the second identity, the integral domain was split into two parts $(0,\epsilon)$ and $(\epsilon,\infty)$ where $\epsilon$ is called the Ewald parameter.
The second term in the final expression can be understood as the short-range part of the Coulomb interactions, whereas the first term deals with the long-range part in the Fourier space. 
However, this expression is divergent because of the term $\left ( \frac 1 {|\vec G|}\right )_{\vec G =0}$, which can be remedied by adding the contribution due to the compensating uniform positive background charges:
\begin{align}
    E_B = -\int d^2\vec r \frac{1/s}{|\vec r|}= - \frac{2\pi}{s} \left (\frac{1}{k} \right)_{\vec k =0}.
\end{align}
Therefore, the sum of the two contributions gives 
\begin{align}
    V_{\rm pair}(\vec r_i -\vec r_j) = \sum_{l} \frac{1}{|\vec r - \vec R_l|}  - \frac{2\pi}{s} \left (\frac{1}{G} \right)_{\vec G =0}.
\end{align}
By choosing an appropriate Ewald parameter $\epsilon = \sqrt{ \pi/ s}$ and by using the fact that $\vec G $ are obtained from $\vec R_l$ via $\frac \pi 2$ rotation followed by $ {2\pi}/{s}$ scaling , it is possible to express $V_{\rm pair}(\vec r_i - \vec r_j)$ in terms of the summation over the lattice vectors $\vec R_l$:
\begin{align}
    V_{\rm pair}(\vec r_i - \vec r_j) =& \sum_{l\neq 0}\left \{ \frac 1 {|R_l|} {\rm erfc}\left( \sqrt{\frac{\pi}{s}}|\vec R_l|\right) \cos\left [\frac{2\pi}{s}(R_l^x r^y + R_l^yr^x) \right ] + \frac 1{|\vec r - \vec R_l|} {\rm erfc} \left (\sqrt{\frac \pi s}|\vec r - \vec R_l| \right) \right \} + \nonumber \\
    &\frac 1 {|\vec r|}{\rm erfc}\left ( \sqrt{\frac \pi s}|\vec r| \right) - \frac 2 {\sqrt s}.
\end{align}
Here, the superscripts $x $ and $y$ on $R_l$ and $r$ denote the $x$ and $y$ coordinates of the corresponding vectors.
Also, the interactions between the electron at the position $\vec r_i$ and its image charges must be dealt with separately:
\begin{align}
    V_{\rm self} &= \sum_{l \neq 0} \frac 1 {|\vec R_l|} - \frac{2\pi}{s}\left ( \frac 1 {|\vec G|} \right)_{\vec G =0 } = \lim_{r \to 0} (V(\vec r ) - \frac 1 {|r|}) \nonumber \\
    &=-\frac{4}{\sqrt s}+ \sum_{l\neq 0} \frac 2 {|R_l|}{\rm erfc} \left ( \sqrt{\frac{\pi}{s}}|\vec R_l| \right ).
\end{align}
Finally, the total Coulomb energy (the sum of interaction energies between $\vec r_i$ and $\vec r_j$ along with its periodic images) is
\begin{align}
    V(\{\vec r_i \}) = \sum_{i<j}^M V_{\rm pair}(\vec r_i - \vec r_j) + \frac M 2 V_{\rm self}.
\end{align}

\section{3. Exchange and hopping processes involving an interstitial}

\begin{suppfigure}[t]
    \centering
    \includegraphics[width = 0.9\textwidth]{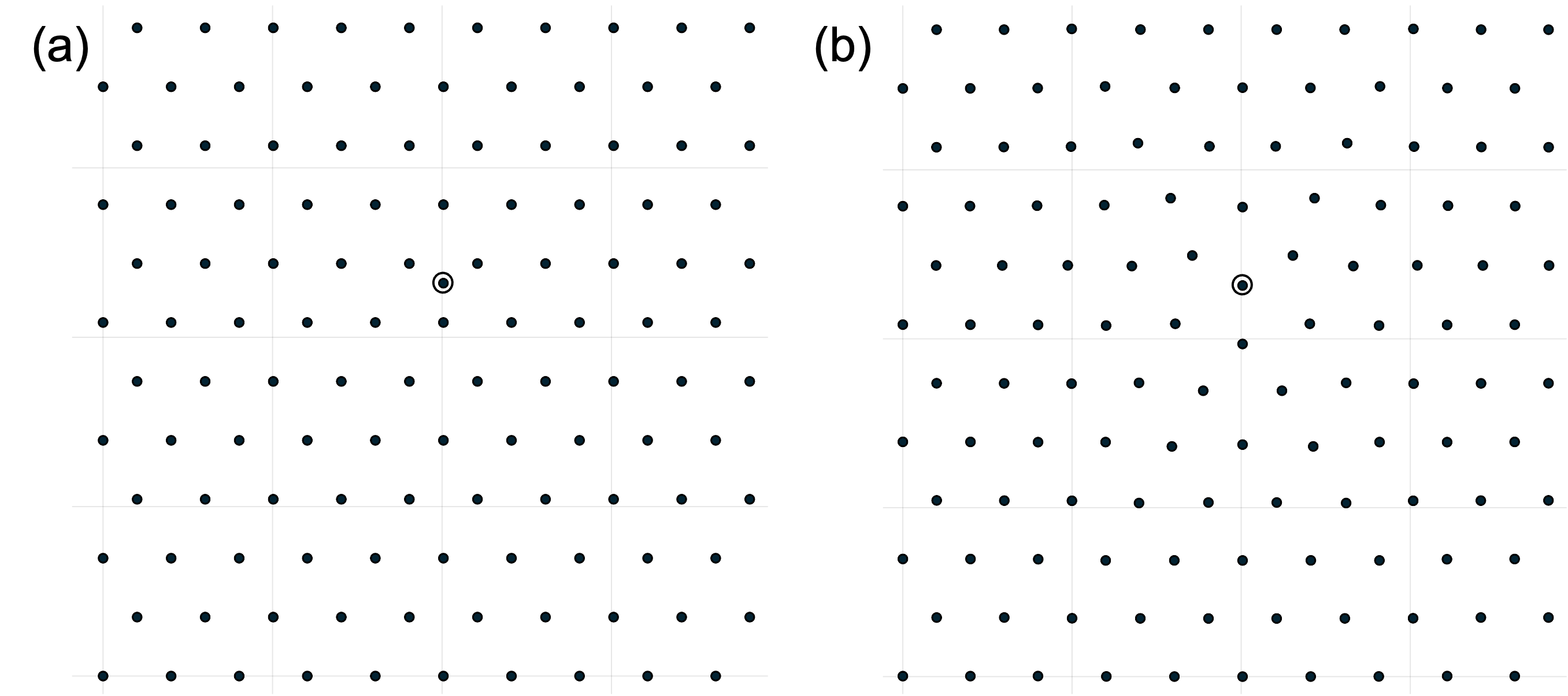}
    \caption{(a) The centered interstitial before relaxation. (b) The interstitial configuration after relaxation. 
    The interstitial sites is emphasized by a circle around them.}
    \label{fig:interstitial configuration}
\end{suppfigure}

We first obtained the relaxed interstitial configuration as shown in Fig. S\ref{fig:interstitial configuration}. 
The energy of an interstitial in the WC with the lattice constant 1 for the system size $M=10 \times 12 +1$ is  0.13555, in agreement with Ref.~\cite{Cockayne1991defect}.

We have calculated 10 different tunneling processes involving an interstitial, as shown in Fig. 1 (b \& c) of the main text. 
For the exchange processes, the initial and the final configurations are identical, ${\bf X_i}= {\bf X_f}$.
In minimizing the action Eq. \ref{eq:classical action}, we discretize the path into 7 points and use the standard trapezoidal method to evaluate the integral, and allow up to $30$ electrons to adjust their positions in the exchange paths.
All the other electron positions are fixed in the initial interstitial configuration.
We used the limited-memory Broyden–Fletcher–Goldfarb–Shanno (L-BFGS) algorithm for the minimization \cite{optim}.
For the hopping processes, the final interstitial configuration ${\bf X_f}$ is obtained by a suitable translation or reflection of the initial configuration ${\bf X_i}.$
Now that ${\bf X_{i}}$ and ${\bf X_{f}}$ are different, we start from the trial path that linearly interpolates the two ${\bf X}_k = \frac k m {\bf X_{i}} + \frac {m-k} m {\bf X_{f}}$ where $m=8$ and $k = 0, 1, ... ,8$, and again allow up to 30 electrons to adjust their positions away from the trial paths.

\begin{supptable}[b]
    \centering
    \includegraphics[width = 0.6\textwidth]{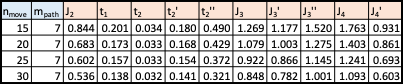}
    \caption{Results of the calculations of $S_a$ for 10 processes as summarized in Fig. 1 (b \& c) of the main text. $n_{\rm move}$ denotes the number of electrons that are allowed to adjust their positions away from the trial path. $m_{\rm path}$ is the number of intermediate points in the discretization of the path integral.}
    \label{table:results}
\end{supptable}

In Table. S \ref{table:results}, we summarize the results of the calculation for 4 different parameters $n_{\rm move}$.
We checked that the results do not change qualitatively for larger values of $n_{\rm move } $ and $m_{\rm path}$. The results summarized in the Fig. 1 (b \& c) of the main text are the ones with $n_{\rm move }= 30 $ and $m_{\rm path} =7$.
In Figures S2 \& S3, we show the tunneling processes involving an interstitial for $n_{\rm move} = 30$ and $m_{\rm path } = 7$. 

Finally, in Fig. S 2, we show the WC sites (red) that are connected from the initial interstitial site (a circled black dot) through a $t_2$ hopping term.
The black WC sites are not connected by $t_2$ term from the the initial interstitial site. 
Therefore, the ground state manifold of the Hamiltonian,
    $H_{\mathrm{eff}} =  -t_2 \sum_{\substack{(n,j,n^\prime) \\ \in (t_2 \: \mathrm{path})}} 
    \sum_{\sigma,\sigma' = \uparrow,\downarrow} 
    c^{\dagger}_{n,\sigma'} f^{\dagger}_{j,\sigma} f_{j,\sigma'} c_{n^\prime,\sigma},$
    in the presence of a single interstitial consists of fully polarized red WC sites and an interstitial, with the spin states of the black WC sites remaining arbitrary.

\begin{suppfigure}[t]
\begin{minipage}[c]{0.4\textwidth}
\includegraphics[width = \textwidth]{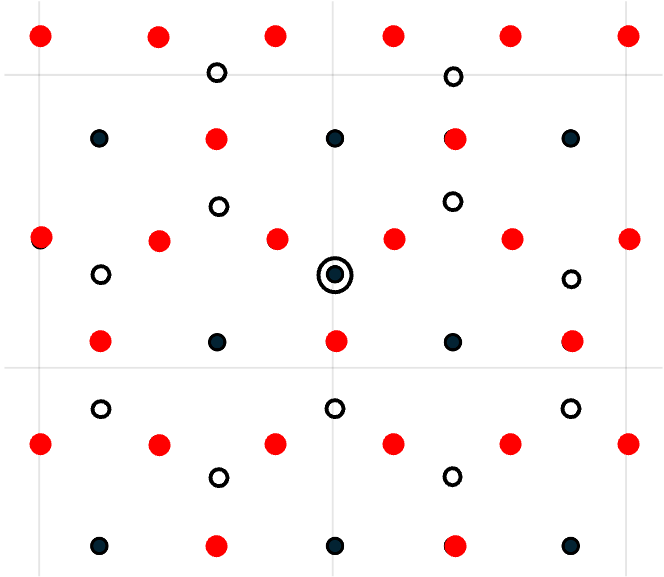}
\end{minipage}
\begin{minipage}[c]{0.4\textwidth}
\caption{The sites connected by the $t_2$ hopping term defined in the main text.
A black dot surrounded by a circle is an initial triangle-centered interstitial.
Other solid dots (black and red) are WC sites.
Red WC sites can be reached by the initial interstitial by repeated applications of the $t_2$ hopping term, whereas black WC sites cannot be.
Empty circles are the interstitial sites that can be reached from the initial interstitial site.
}
\end{minipage}
\label{fig:connectivity}
\end{suppfigure}

\begin{suppfigure}[b]
    \centering
    \includegraphics[width = \textwidth]{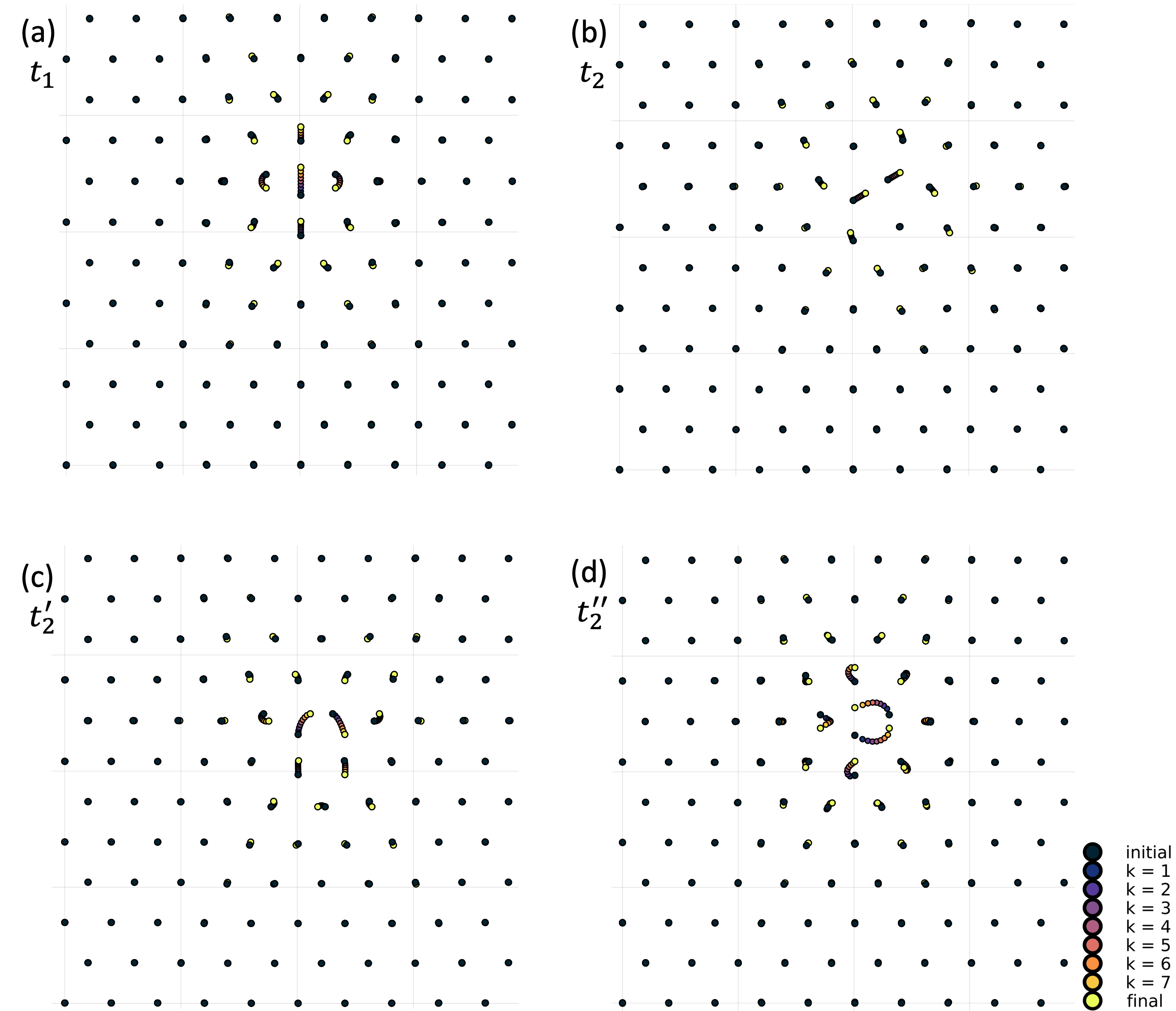}
    \caption{(a-d) The interstitial hopping processes for $t_1,t_2, t_2'$ and $t_2''$, respectively. 
    The corresponding classical actions are $S_{1,{\rm hop}}=0.138$, $S_{2,{\rm hop}}=0.032$, $S_{2,{\rm hop}}'=0.141$ and $S_{2,{\rm hop}}''=0.321$, respectively. 
    The color scheme indicates seven intermediate configurations indexed by k along with the initial and the final configuration.}
    \label{fig:interstitial hopping}
\end{suppfigure}

\begin{suppfigure}[t]
    \centering
    \includegraphics[width = 0.95 \textwidth]{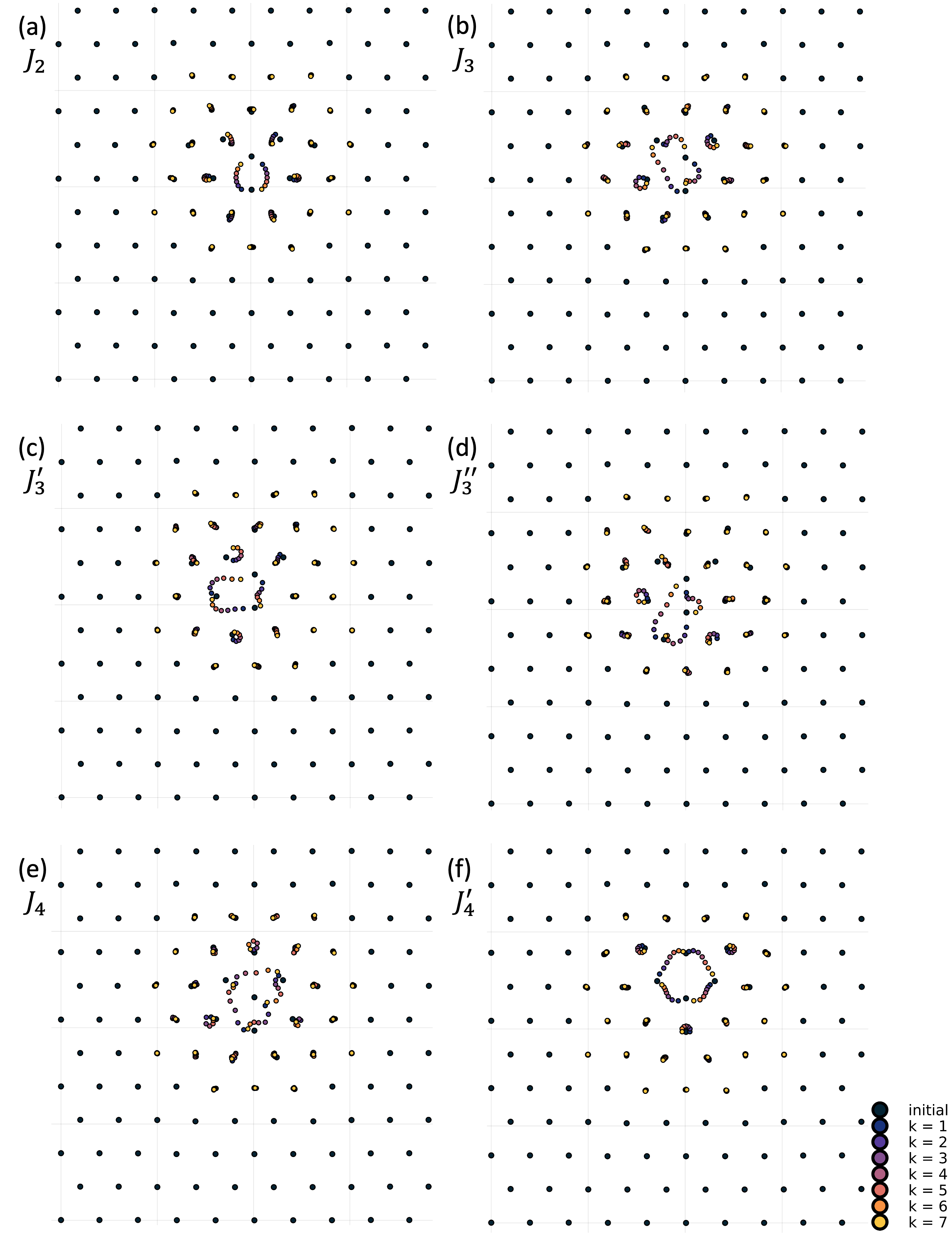}
    \caption{(a-f) The interstitial exchange processes for $J_2, J_3, J_3', J_3'', J_4$ and $J_4'$, respectively. 
    The corresponding classical actions are $S_{2,{\rm ex}}=0.536$, $S_{3,{\rm ex}}=0.848$, $S'_{3,{\rm ex}}=0.782$, $S''_{3,{\rm ex}}=1.001$, $S_{4,{\rm ex}}=1.093$ and $S'_{4,{\rm ex}}=0.603$, respectively. 
    The color scheme indicates seven intermediate configurations indexed by k along with the initial configuration. The final configuration is the same as the initial configuration.}
    \label{fig:interstitial exchange}
\end{suppfigure}

\clearpage
\bibliography{sup.bib}